# *Duration of exposure to inheritance law in India: Examining the heterogeneous effects on empowerment*

*Shreya Biswas*

*Upasak Das*

*Prasenjit Sarkhel*

*Shreya Biswas is an Assistant Professor at the Department of Economics and Finance, BITS Pilani, Hyderabad campus (shreya@hyderabad.bits-pilani.ac.in)*

*Upasak Das is a Presidential Fellow (equivalent to Assistant Professor) of Development Economics in the Global Development Institute, University of Manchester and an affiliate of the the Center for Social Norms and Behavior Dynamics, University of Pennsylvania (upasak.das@manchester.ac.uk)*

*Prasenjit Sarkhel is an Associate Professor in the Department of Economics, University of Kalyani (prasenjitsarkhel@gmail.com)*

# *Duration of exposure to inheritance law in India: Assessing the heterogeneous effects on empowerment*

*Higher duration of programs that involve legal protection may entail gradual positive changes in social norms that can be leveraged by potential beneficiaries in their favor. This paper examines the heterogeneous impact of the duration of exposure to gender-neutral reforms in the inheritance law in India on two latent domains of women empowerment: intrinsic, which pertains to expansion of agency and instrumental which relates to ability to make decisions. The time lag between the year of the amendment in the respective states and the year of marriage generate exogenous variation in reform exposure across women. The findings indicate a significant non-linear increase in the instrumental as well as intrinsic empowerment. Importantly, improvements in education along with increase in the age of marriage and changes in family structure are found to be the potential channels that signal gradual relaxation of social norms and explain the higher returns to exposure on empowerment.*

1. **Introduction**

Legal reforms to ensure gender rights are often at loggerheads with existing social norms. In a review of inheritance law and land rights for women in Sub-Saharan Africa, Hallward-Driemeier and Hasan (2012) note that social norms change slowly in response to de jure reforms and reform benefits take time to materialize. If the *de jure* reforms target social practices that are path dependent[1], it is possible that the social norms get diffused following an S-shaped curve: the rate of change in the social norms start out slow, accelerates as the education and awareness increases and again slow down as the changed norms become widespread (Mackie et al., 2015). Thus, the time duration of the legal reform targeting the female might be crucial for estimates of treatment effect, especially in the context of a traditional patriarchal society with entrenched social norms that work to limit the ability of women to access her rights.

There is compelling evidence in the contemporary literature that securing women's right over land and other productive resources have a positive influence on household welfare (Meinzen-Dick et.al., 2019). To make these rights enforceable, more than 115 countries across the globe have evolved legal provisions for recognizing women's equal property and inheritance[2]. However, the extent to which such legal provisions can have a positive impact on women in terms of their rights can potentially be time dependent especially in the context where social norms are rigid and work against women. This is aggravated by the fact that because of low levels of awareness and social concerns, the beneficiary females are often unable to register their claims and appeal in the court of law. Therefore, the existing social practices that have traditionally discriminated against women might take time to adjust to the legal provisions and hence the program effects are likely to manifest themselves after a time lag, thereby enhancing the significance of the duration of exposure to the program. Without the associated positive changes in norms, the program effects may actually decay. In this paper, we use the program duration of the Hindu Succession Amendment Act (HSAA), a significant legal reform passed in parts of India for ensuring equitable female property right, to investigate the heterogeneous impact on women empowerment and identify the possible channels through which the reform worked.

---

[1] Following North (2005) path dependence is defined as the institutional structure obtained from the past that embodies a set of belief that are resistant to change proposed by legal reforms. As it is, path dependence can arise from either increasing returns from the status-quo or high switching cost.

[2] UN Entity for Gender Equality and the Empowerment of Women (UNWOMEN), Realizing Women's Rights to Land and other Productive Resources, 2013, HR/PUB/13/04

The HSAA recognized the daughters as coparceners, allowing them to ask for partition of property occupied by the intestate family. This was an amendment of the previous Hindu Succession Act of 1956 that governed the inheritance rights pertaining to ancestral and self-acquired property of Hindu, Sikhs, Jains and Buddhists in India (Agarwal, 1998). The impact of HSAA on the economic and social status of women in India has attracted substantial research in the last decade. While most of these studies have extolled the virtue of the Act that secured property rights of women in enhancing their bargaining power (Deininger et al.,2018), increased autonomy (Mookerjee, 2019) and higher participation in the labour market (Heath and Tan, 2020), the long term effects of the program in terms of the duration of the exposure are not yet explored. As argued by Agarwal et al. (2020), just enacting a law might not be sufficient in altering household and individual behavior and attitudes. This is especially true in the context of India with entrenched patrilineal norms, where parental resistance to sharing the immovable property with daughters is well documented (Agarwal, 1994; Chowdhury, 2017). Further, Agarwal et al. (2020) also point out the gap between reforming the law and awareness. We argue as the program duration increases over time, the associated social norms may start altering along with an increase in familiarity of the potential beneficiaries with their legal rights and accordingly the duration effects via increased learning might reinforce the positive effect on female autonomy and other well-being indicators including education among others. Though literature has assessed the impact of the reforms on a wide set of indicators of female empowerment through an average effect over time, much less clear is the extent of the impact on the duration of the program exposure that may potentially capture the learning effects along with the patriarchal social norms becoming less stringent. In this paper, we explicitly account for the duration of exposure to HSAA and estimate its heterogeneous impact on the intrinsic and instrumental/functional empowerment of the married Hindu female across Indian states.

Our work stands distinct from others in two significant ways: firstly, most of the studies that dealt with program exposure have used binary indicators through exploiting state and birth cohort variation that, however, doesn't explicitly account for the duration of HSAA. As discussed by King and Behrman (2009) this could result in omissions of important effects like delays in program implementation and familiarity with the implemented provisions. From the point of view of the female beneficiaries, the law remains extraneous until they acquire the capacity to overcome the social sanctions associated with claiming the parental property in the court of law. More importantly, there might be significant heterogeneity in the ability of the female claimants to understand their legal rights and they might be situated differently in the learning curve given their social status (Behrman and King, 2008). On the other hand, if the patriarchal norms remain rigid over time, the impact may remain constant or even start diminishing in some contexts. Our paper defines exposure as the time difference between the year of marriage and the year of reform that is likely to capture the learning effects or changing social norms due to program exposure more succinctly on marriage as well as post-marriage familial outcomes and assess whether there is an increasing return to exposure on empowerment.

The second part where we contribute to this growing literature is to underscore the importance of two prominent latent domains of female empowerment or agency that firstly include externally observed features like the ability to make decisions as well as internal thinking (Kabeer, 1999; Mosedale, 2005; Richardson, 2018). The indicators in the first domain are largely means to attain other development goals and are commonly known as instrumental empowerment measures/ indicators. The second domain largely pertains to expansion of agency and an individual's ability to make strategic life choices. Here empowerment is seen in the light of the capability approach propounded by Sen (1985) and commonly known as intrinsic measures/ indicators.

Since women's social status and socio-economic outcomes are often tied to their financial worth, exposure to property rights can have a positive impact on female agency, but arguably may have different gains on the instrumental and intrinsic empowerment indicators. Through the wealth effect created by recognizing the daughters as coparceners and giving them legal rights to ancestral property, their post-marriage bargaining power within the household may improve. This can potentially boost their ability to take decisions and make effective choices. With higher exposure to the property rights laws, the awareness levels as well as patrilineal social norms related to sharing of property rights with daughter may get relaxed. Hence we hypothesize that exposure to the reforms would enhance female empowerment with respect to both intrinsic as well as instrumental autonomy. However, the extent to which duration would impact upon these two dimensions of empowerment is an empirical question and our study examines this aspect as well.

Prior to the amendment that was implemented across India in 2005, the implementation of HSAA took place in five states in earlier period. The Indian state of Kerala did it in 1976, followed by the undivided Andhra Pradesh in 1986, then Tamil Nadu in 1989 followed by Maharashtra and Karnataka in 1994. We use this state-wise staggered variation in implementation and also the year of marriage to calculate the time lag in years between the marriage and amendment of the Act. Considering timing of marriage to be exogenous to the passage of HSAA, we utilize this exogenous variation in the time lag and term it as *exposure* to the amendment to examine its potential effects on post-marriage female empowerment indicators. Our data comes from the third round of National Family Health Survey (NFHS-3) that is a nationally representative survey conducted in 2005-06. The information on women decision making and other autonomy indicators recorded in the survey also enables us to construct indicators of both intrinsic and instrumental empowerment that we compare across the respondents from the reform and non-reform states. Our results suggest that a longer duration of exposure to HSAA has a relatively larger effect on the instrumental part of empowerment. However, we find significant heterogeneity in the impact as the gain in instrumental empowerment cannot be confirmed for eligible women belonging to scheduled tribes. We also find that one of the main channels through which the positive empowerment effect of HSAA operates is through education and thus, our results underscores the need for affirmative strategies for human capital augmentation, targeting disadvantaged caste groups along with strengthening the property right reforms. In addition, we identify secular gains from higher age at marriage and being in a nuclear family that comes through the HSAA especially on intrinsic and instrument empowerment indicators respectively indicating a definite change in the existing patrilineal social norms

The remainder of the paper is organized as follows. Section 2 provides the legal and institutional framework of the study. Section 3 describes the data and Section 4 outlines the identification strategy of our study. Section 5 provides the summary statistics while Section 6 highlights the results and Section 7 provides a discussion and concludes.

## 2. Hindu Succession Act Amendment, 1956

The Hindu Succession Act, 1956 (HSA), governed the inheritance rights pertaining to ancestral and self-acquired property of Hindu, Sikhs, Jains and Buddhists in India. There are two schools of law governing the inheritance practices among the Hindu – Mitakshara and Dayabhaga. The Mitakshara school of Hindu law prevailing in most of India distinguishes between self-acquired property and ancestral or joint property. The self-acquired property refers to tangible assets like land, house, or jewelry that the Hindu male buys with his own funds. The self-acquired property of a Hindu male household is inherited by his sons, daughters and widowed wife (class I heirs) in the event of the death of the household head intestate and is shared equally among them. On the other hand, the ancestral property inheritance is based on the system of 'coparcenary' consisting of males heirs belonging to three generations. Only the male heirs have the birthright over ancestral property and not the daughters. However, in the event of the death of the household head, the daughters and wife along with the son will also have equal rights on the notional portion of the ancestral property of the deceased head. The right to inheritance (or lack of it) distorted the inter-generational wealth transfer in favor of sons and against the daughters of the households[3].

The Hindu Succession Amendment Act, 2005 (HSAA, 2005), rectified the HSA recognizing the daughters as coparceners and also allowing daughters to ask for partition of property occupied by the intestate family. Five states in India amended the HSA related to the coparcenary rights prior to the amendment in 2005. The Kerala Joint Family System (Abolition) Act which came into force in 1976 abolished the joint family system in Kerela. Other four states also amended the law: Andhra Pradesh (1986), Tamil Nadu (1989), Maharashtra (1994) and Karnataka (1994). These four Indian states amended the HSA to include daughters as coparceners. The amendment stated that all unmarried women at the time of passing the law would inherit an equal share of ancestral and joint property and will be included in the coparcenary system, thus excluding the married daughters from the coparcenary system.

The inheritance rights of daughters in India have become more gender-neutral in past few decades, however lack of awareness among women regarding their land rights and social norms remain primary obstacles to actual land ownership by women in India (Kelkar, 2014), and land remains a critical asset that is inherited. Social norms act as a hindrance in the ownership of assets by women in several economies of Asia and Africa, and gender-equal legislative amendments can only slowly change such sticky norms over time (Munoz-Boudet et al., 2012). The study by Sircar and Pal (2015) find differences in awareness level in Bihar (a state where no amendment to HSA was passed before 2005) and that in Andhra Pradesh (where HSA was amended in 1986). In Bihar, only 42% of the surveyed women were aware of inheritance rights compared to 57% in Andhra Pradesh. Further, in Andhra Pradesh 30% of the women surveyed inherited land as opposed to only 8% in Bihar. These exploratory findings suggest that awareness and social norms could improve slowly over time and regulatory changes though necessary, may not alone be sufficient to increase land rights of women, particularly in a society that has a strong patriarchal lineage and is inherently gender-biased.

---

[3] If a household consists of father (generation I), mother, two daughters and two sons (generation II). In case the father (generation I) dies intestate leaving joint property worth Rs.12,00,000, the wife and two daughters will inherit Rs.80,000 each (1/5th of the notional share of 1/3rd of the joint family). The two sons will inherit Rs.400,000 (1/3rd of the ancestral property) and Rs.80,000 (1/5th of the notional share of 1/3rd of the joint family) each.

## 3. Data and variables

We use data from the third round of the NFHS-3 conducted in 2005–06. It is a large scale nationally representative survey of households across all the then 29 states of India. The survey administers basic information about all the members of the sampled household along with the common socioeconomic data on household characteristics like caste, religion, wealth status, and ownership of durable assets. Further, it also gathers information on year of marriage and state of residence for all women in the household between the age of 15 to 49 years apart from their education status, labour participation, husband's occupation and education among others. One unique feature of the survey is it collects extensively information on the status of autonomy of the female members (15 to 49 years of age) in the sampled households across different dimensions of family relations. For example, the survey asks about her decision making ability about healthcare, major and minor purchases within the household, how husband treats wife in front of others, wife's perception regarding domestic violence among others.

As discussed in this paper, we categorize our main outcome variable, female empowerment into two prominent latent domains or schools of thought. In the first group, those indicators, which in some way are means to attain other development goals can be defined as the instrumental empowerment ones. In other words, instrumental empowerment indicators are those which focus on choices (Narayan-Parker 2002; Alsop and Heinsohn 2005; Petesch et al. 2005; Alsop et al. 2006; Ahmed and Hyndman-Rizk 2020). This refers to the enhancement of power or agency through interaction of two building blocks: an individual's or a group's capacity to make purposeful choices within the operational structure that includes the broader political, social, and institutional context of informal along with formal rules and norms, within the purview of which individuals pursue their interests. Individual capabilities such as education, health, societal position, taking financial decisions or ability to think of a better future are determined by these individual assets or institutions or both.

The second group of scholars views empowerment in the light of the capability approach propounded by Sen (1985). Empowerment here pertains to expansion of agency and the expansion in an individual's capacity to make strategic life choices in an environment where this capacity was previously denied to her (Kabeer, 2001; Ibrahim and Alkire, 2007). As argued in Kabeer (2001), expansion and improvement in social constraints like social norms that lead to deprivation and neglect among the females should come with the purview of intrinsic empowerment. Because women should be active agent in the process of development, improvements in entitlements should include better education, mobility, freedom from gender-based violence and control over sexual behavior among others (Sen, 1999).

We broadly follow the framework discussed in Malapit et al. (2019) to construct the intrinsic and instrumental empowerment measures. The set of questions covering attitude of women towards intimate partner violence, how husband treats wife, respect and trust in marital relation constitute the indicators considered under intrinsic empowerment measure. Instrumental empowerment considers the indicators related to women's freedom of movement, financial autonomy, and the ability to make household decisions. Table 1 enumerates the questions considered under each dimension of empowerment. The intrinsic measure is the sum of the z-scores of indicators considered for the domain. Similarly, the instrumental measure is the sum of the z-scores of indicators mentioned in the table. The overall empowerment measure is the sum of intrinsic and instrumental measures. Similar to Heath and Tan (2020), our empowerment variables are the standardized values for the instrinsic, instrumental and overall empowerment indicators.

<<Insert Table 1 here>>

In the regression specification, we control for potential individual and household level confounders including age of the woman, caste to which the household belongs, place of residence, and wealth index. Further, we control for year of marriage fixed effects and state fixed effects.

4. **Empirical Strategy**

Our identification strategy exploits the exogenous variation in exposure to HSAA for Hindu women in the reform states who were married after the implementation of the amendment in their respective states. This exposure is measured in terms of the number of years between her marriage and the implementation of the amendment. Accordingly, the exposure for women who were married in the non-reform state remains zero.

Let $y_{ist}$ is the outcome variable as discussed in the earlier section, which is a measure of instrumental or intrinsic empowerment of woman, $i$ from state, $s$ married in the year $t$. Let $reform_i$ be a dummy variable that takes the value of 1 if woman, $i$ belongs to a state where HSAA has been implemented (Kerala, Andhra Pradesh, Maharashtra, Tamil Nadu and Karnataka) and $Z_{is}$ is binary variable that takes the value of 1 if woman, $i$ got married after the amendment of the HSA in their respective state, $s$. Please note that this variable is 0 for all women from the reformed states, who got married before the amendment and also for those who belong to non-reformed state irrespective of when they got married. Therefore, for a woman from a reformed state who got married after the amendment, $reform_i * Z_{is}$ takes the value of 1 and 0 for everyone else. Let $exposure_{ist}$ be the difference in years between the marriage and the passage of the amendment for a woman, $i$ from state, $s$. Note that this variable would be zero for all those women, who belong to states which did not pass the amendment or those from the reformed state who married before the passage of the amendment. To estimate the causal effect of the exposure of the amendment of outcome variables, we estimate the following model:

$$y_{ist} = \alpha + \beta_1 exposure_{ist} + \beta_2 reform_i * Z_{is} + \beta_3 reform_i + \beta_4 X_{ist} + \lambda_s + \delta_t + \varepsilon_{ist} \quad (1)$$

Here $X_{ist}$ is the vector of individual and household level characteristics that are potential confounders to the outcome variable. $\lambda_s$ are the state level fixed effects that would control for all confounding variables that vary across states and $\delta_t$ are the year of marriage fixed effects that would control for idiosyncratic year specific shocks/ changes that can affect the outcome variables. Here $\beta_1$ is the coefficient of interest that captures the causal marginal effect of exposure to the HSAA, controlling for secular effects of marriage after amendment and other potential confounders. Because this amendment was implemented for Hindu woman, we ran this regression for them only.

In our context, arguably the variable $exposure_{ist}$ as we define would account for the exposure to the reforms that get manifested for a woman during the time of marriage. Since we take the time lag between the passage of the amendment and year of marriage within that reformed state, we argue higher this time lag is, lesser would be the resistance from pre-existing social practices towards execution of the reforms. Higher duration may also imply better information about the legal covenants.[4] Notably, in our specification, we are controlling for year of marriage fixed effects and the age of the wife during the time of the survey. These variables would capture the secular effect on empowerment due to time since marriage and also any unprecedented year of marriage shocks that may systematically affect female agency. In addition, we are also controlling for the effect of the passage of HSAA through $reform_i * Z_{is}$ as done in most of the other studies on the reforms. So, we argue $\beta_1$ would give us an unbiased causal estimate for exposure to HSAA after adjusting for the secular effects of the introduction of HSAA, year of marriage and age of the woman along with other possible confounders.

A pressing concern with the aforementioned difference-in-difference specifications is that there may be time-varying omitted variables specific to states that becomes correlated with the passage of the reforms. In other words, the identifying assumption here is that the decision of marriage is not correlated with the introduction of the reforms. As argued by Mookerjee (2019), we can allay these endogeneity concerns that arise due to selection in the timing of the marriage since these state amendments were often passed on the retrospective.[5] We also argue that endogeneity concerns in exposure to the amendment are minimal since parents are highly unlikely to delay the marriage of their daughters because of higher exposure to the program as defined by us. If the parents or daughters are already aware of the provisions of the amendment, there are no reasons to believe they would delay their marriage just because of increasing the number of years between her marriage and the amendment. On the other hand, if the parents or daughters are unaware of the provisions, the decision of marriage would then be purely exogenous to the amendment year. So concerns surrounding endogeneity due to self-selection of the timing of marriage through increasing the exposure to the program is highly unlikely.

Endogeneity concerns also arise through the unobservables which are correlated because of self-selection of families through migration or residential relocation (Stark and Taylor, 1991). Because of the passage of HSA, households looking for potential brides may move to the reform state because they can take advantage of the implementation of HSAA. This non-random sorting can potentially bias our estimates through the omitted variables raising questions about causality through our empirical framework. However, residential relocation is unlikely to be the cause of concern for our case, as the literature indicates spatial mobility at the household level to be extremely low in India (Munshi and Rosenzweig, 2009; Munshi, 2016; Rowchowdhury, 2019). In addition, interstate migration especially during the period we examine is found to be very low (Premi, 1990; Singh, 1998).

---

[4] Of course, with time as argued earlier, if the social norms remain rigid, we might as well observe perverse effects on empowerment indicators.
[5] As argued by Mookerjee (2019), the amendment in Andhra Pradesh was formally passed on May, 1986 but had come to effect from September, 1985. Similarly, the amendment started retrospectively from June, 1994 though the act was officially passed in December 1994.

5. **Summary statistics**

Figure 1 presents the age-cohort wise summary statistics of the empowerment measures for the married Hindu women in our sample. We find that the average empowerment measures for married women till 30 years is below the average for older cohort in both reform and non-reform states. This is in line with the extant literature that indicates that younger wives are less empowered than older women in India (Desai and Andrist, 2010). However, for both the age-cohorts, the individual instrumental and intrinsic measures are higher for women from reform states compared to women from non-reform states. Figure 2 presents a local polynomial graph to show the relationship between the empowerment indicators and duration of exposure for all married women from the reform states. While no apparent association is observed for instrumental and overall empowerment, we find a secular rise in intrinsic empowerment across the distribution of exposure.

<<Insert Figures 1 and 2 here>>

The summary statistics of other characteristics of women are given in Table 2. The top panel provides the summary for the overall sample along with treatment and control groups. The age at marriage and years of schooling is higher in reform than non-reform states. Also, within reform states, the outcomes improve for women who got married after the HSAA (treatment group) compared to the women who were already married at the time of HSAA. The bottom panel provides the exposure wise summary for the treated sample. Age at marriage and years of schooling increases with exposure and inter-spousal educational difference reduces with exposure.

<<Insert Table 2 here>>

6. **Main Results**

This section presents the regression results to estimate the effect of exposure to HSAA on the z-score of women empowerment measures. The results are given in Table 3. The coefficient corresponding to the *exposure* variable from equation 1 gives us the causal impact. Column 3 indicates that each additional year of exposure increases the women's autonomy by 0.012 standard deviations. For example, ceteris paribus, a treated woman in reform state having one year of exposure to HSAA is found to be 0.012 standard deviations more empowered on average than the woman from a non-reform state. Columns 1 and 2 present the results for instrumental and intrinsic empowerment, respectively. The estimates show that each additional year of exposure has a larger effect on instrumental empowerment (0.012 standard deviations), while the effect on intrinsic empowerment is found to be lower (0.006 standard deviations). Nevertheless, both these variables are found to be statistically significant at 1% level of significance.

<<Insert Table 3 here>>

We also obtained the causal estimation of the impact on the individual measures of instrumental and intrinsic empowerment. On the former, we find that with each additional year of exposure, the marginal effect in the probability of women having a say in household purchases (both small and large), say in health, say in how to spend husband's earning, or freedom of mobility (visit to family, clinic, and market) increases in the range of 0.2 to 0.6 percentage points. However, we do not find evidence of any causal association of exposure with the probability of women owning an account, having their own money, and deciding on how to spend their own money.[6]

The findings on the indicators of intrinsic empowerment reveal higher exposure to the reform for an average woman is significantly related to a higher probability of her husband showing respect towards her. For example, an additional year of exposure to HSAA increases the likelihood of the husband not accusing the wife of being unfaithful, trusting with money, and allowing her to meet friends by about 0.2% each though no significant effect on the perception towards domestic violence is observed[7].

Notably, the positive implications of women empowerment are qualitatively similar to other studies on HSAA (Mookerjee, 2019; Heath and Tan, 2020). However, our findings additionally indicate that there is a long term effect of the policy whereby the impact on women who are eligible to take benefits of the policy is not uniform but increases as the exposure to the reform increases. Likewise, it is possible that the effects of HSAA are non-linear to its exposure in the sense that women who are exposed more to the reforms get disproportionately more benefits than those who are less exposed. To test this, we examine whether the long-run impact of exposure on women empowerment is significantly different than the impact in the short run. This *exposure level* variable takes the value one for exposure up to five years, equals two for six to ten years of exposure, and then takes the value three for any exposure greater than ten years. Initially, for the first ten years of exposure to HSAA, the effect seems to be modest; however long term impact is found to be much stronger (Figure 3).

---

[6] The results from the regression are given in Appendix Table A1.
[7] The results from the regression are given in Appendix Tables A2 and A3.

<<Insert Figure 3 here>>

To sum up, we find that higher exposure to HSAA is found to have a significant positive effect on women empowerment. If we group the empowerment indicators into the instrumental and intrinsic categories, our findings indicate that the marginal effects of an additional year of exposure is higher for instrumental empowerment in comparison to the intrinsic measure though positive impact on both are observed. Further, a heterogeneous increasing non-linear relationship between exposure and empowerment is found whereby the gains for women who are more exposed are found to be disproportionately higher.

**6.1 Falsification test**

The variation in the year of implementation of HSAA across states enables us to run placebo tests with different years of the reform. These regressions are likely to yield insignificant estimates for the causal impact of the placebo exposure to the reform if our hypothesis of higher empowerment through HSA exposure holds true. Accordingly, first, we consider 1970 as a randomly selected reform year for all the five states that implemented the HSAA.[8] Kerala was the first state to pass the reform in 1976; hence, for this analysis, we consider the sample of Hindu women who were married before 1976. The control group is the sample of women from non-reform states along with those from the reform states who were married before 1970. The *exposure* variable here is defined as the year of marriage minus 1970 for the treated women in reform states and zero for other women. Columns 1-3 of Table 4, which present the results from this falsification test indicate that the effect on empowerment measures is insignificant for this sub-sample. We repeat the same exercise with another random placebo year (1982).[9] Since the HSAA was passed in Kerala before 1982, all observations from Kerala are dropped for this analysis. Using similar method as above but replacing the year, 1970 with the year 1982, the effect on the variables pertaining to empowerment remains statistically insignificant. The results conform to our hypothesis that exposure to HSAA is related to improvement in the empowerment of treated women.

---

[8] To ensure sample adequacy, we consider the period 1968 to 1975 and through a random draw chose 1970 as the placebo year.

[9] For this we consider the period 1977 to 1985 and through a random draw chose 1982 as the placebo year.

<<Insert Table 4 here>>

Further, we examine the impact of exposure to the reforms on non-Hindu women. If indeed, HSAA is the cause of improvement in the empowerment indicators for Hindu women in the reform state, similar regressions should potentially yield insignificant effect on non-Hindu women since the property right reforms were applicable for only the Hindu women and there was no change in the property laws for non-Hindu women. Table 5 presents the results from the regressions only for the non-Hindu women. As one would expect, the impact of exposure to HSAA is found to be statistically insignificant for instrumental as well as intrinsic empowerment measures. Of note is the fact that these falsification results to some extent potentially ensure that the parallel trend assumption holds in the sense that there was no systematic difference in trends for indicators related to women empowerment in the pre-reform period. Hence it reassures us that the estimated effect is not driven by the confounders that include inherent improvement in women status, education or economic status among others which can be systematically different in the reform states in comparison to the non-reformed states. Rather the improvement in the status of women in the reform state is found to be causally related to the exposure to the reforms.

<<Insert Table 5 here>>

### 6.2 Sub-sample analysis
#### 6.2.1 Rural-urban differences

The gap between rural and urban areas in India across dimensions including poverty, inequality, education, occupation choices and wages is well documented (Deaton and Dreze, 2002; Hnatkovska et al., 2013). Further, the way HSAA might be implemented might be different as well. For example, the enabling institutions and awareness levels about the law may be different and also change systematically over the years potentially favoring the women from urban areas. Because of these possible factors, implications on women empowerment due to HSAA might be different across rural and urban sectors. Accordingly, we examine the possible differential effects of the reforms in rural and urban sector separately. Figure 4 indicates that the exposure to the reforms is found to have a significant effect on empowerment indicators in both the sectors, but the effect size seems to be higher in urban areas. For example, an additional year of HSAA exposure is found to increase the overall empowerment by 0.019 standard deviations in urban areas while in rural areas, this increase is by 0.009 standard deviations. Hence the effect size in urban areas is more than double of that in the rural areas. Further, we find a significant increase in intrinsic empowerment due to HSAA exposure in the urban areas which is not the case in rural areas.

<<Insert Figure 4 here>>

### 6.2.2 Caste groups

The disparity in landholdings, education and consumption across caste groups in India is well documented in literature (Deshpande, 2000; Khamis et. al., 2012; Besley, et. al., 2016). It may be noted that the Scheduled Castes (SC) and Scheduled Tribes (ST) have suffered from severe social exclusion and discrimination from historical times and lag behind the Upper Castes (UC) and the Other Backward Castes (OBC) in the different indicators of welfare. However, when UCs and OBCs are compared, the former is much better off than the latter. Likewise, if land ownership is significantly lower among households belonging to the historically marginalized Scheduled Castes (SC) and Scheduled Tribes (ST), then the HSAA is likely to have a lower impact on treated women from these households compared to the Hindu women from other social groups. Further, marital rituals are also largely different among tribals across India compared to other social groups with the cultural and gender norms being more egalitarian (Maharatna, 2000; Mitra, 2008). Hence gains from HSAA might be dissimilar across caste, and to test this differential effect, we divide our sample of Hindu households into four groups depending on the social groups they belong to: UC, OBC, SC and ST and run the same regressions.

Table 6 presents the results from those regressions. The findings reveal significant positive effect on overall empowerment for the women belonging to the UC, OBC and SC with the effect being largest for SCs. For all these three groups, there is a significant rise in the instrumental empowerment indicator but it is only for the SCs, where we observe significant gains in the intrinsic empowerment index as well. For STs though, we do not observe any such effect whatsoever on these empowerment indicators. There might be multiple reasons behind this. Firstly as already mentioned, studies have indicated that STs are highly disadvantaged economically and fare worse than SCs and Muslims as well (Maharatna and Hawley, 2005; Banerjee and Somanathan, 2007). Hence the likelihood of them having ancestral property that can elevate the position of the women within the household is unlikely. Secondly, since they are marginalized and many of the tribes are not even main-streamed, the chances of them being aware of the developments related to property rights are less, because of which we do not observe any gains from the reforms. Finally, it is possible that because of the already egalitarian gender norms among the tribals, an average ST woman is otherwise better empowered in comparison to a similar woman from UC, OBC or SC community (Mitra, 2008). Hence because of the higher base levels, it may be the case that no significant effect is observed. However our data shows no such pre-reform period difference in empowerment levels for ST women and others. Hence it is likely that the first two reasons would hold but not the last one, though this needs further research.

<<Insert Table 6 here>>

### 6.2.3 Young wives

Studies have highlighted that newly married wives are less empowered than mothers and older wives in the household (Desai and Andrist, 2010; Samari, 2017). In fact, a study across nine countries indicates that young women face substantially higher risk of experiencing Intimate Partner Violence (IPV) in comparison to the relatively older women (Stockl et al., 2014). As pointed out by Desai and Andrist (2010), majority of the newly married brides in India live with her in-laws. The younger brides are "expected to conform to the lifestyle of a new family" and likely to be more "docile". However, exposure to the property rights reforms would also be higher for these young brides because of which their position in the family may enhance. Hence the direction of the impact for the younger married women and the older ones is not exactly clear and remains an empirical question. We study this issue by separating the sample of Hindu married women: those who are married in the year 2000 or after and those married before and running separate regressions for the two groups. Here, it should be noted that for the young wives the control-group is married women from non-reform states only since the young wives in reform states are exposed to HSAA by construction.

Figure 5 presents the coefficient of *exposure* variable, and we find that there is no significant effect for young brides though the gains for older women are found to be substantial in terms of overall empowerment largely driven through improvements in instrumental empowerment indicator. One reason for this might be the fact that the associated property rights are not materialized till death of the bride's parents, hence in the case of young brides, the chances of this is lower while for an older woman, the chances remain higher. Of course, the other part of the story can possibly be the inertia of the patriarchal social norms in the Indian context, where even an egalitarian inheritance right is unable to alter the household dynamics, especially for young brides. This remains another issue that can potentially be explored in future research.

<<Insert Figure 5 here>>

### 6.3 Possible channels and Heterogeneous impact

Having established a positive impact of the exposure to HSAA on women empowerment, we consider improvement in women's age at marriage, changes in husband and wife's education and change in family structure as the potential channels to explore the marginal returns to exposure to HSAA on women empowerment. Notably a society with entrenched patriarchal norms is likely to be resistant in educating females, would prefer their early marriage and also would prefer brides to stay with in-laws post marriage. We argue if patrilineal social norms that start showing positive changes in favor of women, that would get reflected in female education, age of marriage and family structure as well.

#### 6.3.1 Education and women empowerment

Education is widely seen as a key element that raises the status of women and in the process empowering them by expanding their knowledge and skills (Jayaweera, 1997). Women's property rights can have substantial bearing on their empowerment through education. With the implementation of property right reforms, it is possible especially in a patriarchal society that parents may strategically want to avoid compliance with the reforms and not provide inheritance of their property to the females. In this case, they may compensate for this disinheritance by making higher investments in education for the daughters and paying more dowry (Roy, 2015). In other cases, since women would have higher bargaining power in the household because of the property rights, they may demand more investment in education from the parents, especially if the private return to education is high. Hence it is likely that education of the women would improve with the passage of the HSAA. Literature indicates that the reform had a significant impact on improving education for the women (Deininger et. al., 2013, Roy, 2015; Bose and Das, 2017). This improvement in education of the wives can potentially improve post-marital bargaining power for the women which can result in higher empowerment.

Because of this potential increase in education of wives, the average education of the husband may also increase because of prevalence of assortative mating based on education (Mare, 1991; Hahn et al.,2018). Controlling for women's education, increase in husband's education may results in higher empowerment because his education may enable him to respect his wife and enable her to take more decisions of her own. Likewise, it may also happen that an additional year of education may make him take the upper hand in the spousal relationship which may be detrimental to his wife's empowerment. In other words, the direction of the effect of improvements in the husband's education through the HSAA on his wife empowerment indicators may not be clear. Accordingly, in this section we examine if increased education of the husband and the wives because of the passage of HSAA can potentially stand as one of the channels for observing higher empowerment.

For this purpose, we first examine if exposure to HSAA has a significant association in improving educational attainment of the wives and the husbands. As one would expect, our findings indicate an additional year of exposure to HSAA is associated with an average increase in schooling for the wives by 0.003 standard deviations. This also led to an average increase in husband's education by 0.006 standard deviations, which is found to be statistically significant at 1 percent level (Column 2, Table 7). Further, we examine if exposure to property rights reforms led to an increase in educational homogamy given by the z-score of inter-spousal educational difference (the difference between husband's and wife's years of schooling). An increase by 0.003 standard deviations is observed, which indicates there has been a rise in husband's education relative to the wife's, albeit very moderate.

<<Insert Table 7 here>>

Table 8 presents the heterogeneous effect of exposure to HSAA based on education, where we test whether the gains in education of the wife and husband separately because of the property rights reforms resulted in systematic increase in empowerment of the former. The dummy variable, *high_edu* takes the value one for women with positive z-score for educational attainment and zero otherwise. In the regression, the coefficient pertaining to the interaction between *exposure* variable and *high_edu* dummy gives the differential impact of exposure with a more than average educational attainment. Our findings indicate that an increase in an additional year in exposure to HSAA has a 0.006 standard deviations higher impact on intrinsic as well as overall empowerment for highly educated females respectively (Columns 2 and 3), both of which are found to be statistically significant at 5 percent level or less. With respect to husband's education, there is an increase in instrumental empowerment by 0.004 standard deviations only at 10 percent level. However, unlike women's education that has a higher bearing on intrinsic and overall empowerment, husband's education is found to have limited effect only on instrumental indicator. This indicates that female education is important in accentuating intrinsic and instrumental empowerment and serves as one of the channels for observing improvements in overall empowerment through HSAA. The findings emphasize that the role of education in improving the women's bargaining power vis-à-vis the husband remains critical in enhancing intrinsic empowerment in Indian society where social norms are prevalent.

<<Insert Table 8 here>>

### 6.3.2 Age at marriage and women empowerment

Early marriage of women in developing economies is deep-rooted in norms (Caldwell et al., 1983) that inhibit the educational attainment and labor force participation of women (Maertens, 2013). The young women in patriarchal societies do not have the option to voice their opinion regarding marriage timing, and the age of marriage is primarily driven by the society's perceived marriageable age. Studies have documented the delay in marriage for women is related to significant gains in social status in the family post-marriage and reduction in domestic violence and hence can enhance women agency (Jensen and Thornton, 2003, Yount et al., 2018). Of note is the fact is that this effect is observed accounting for the increase in educational level.

To establish age of marriage as a potential secular channel that can enhance women empowerment, we need to test whether there is an associated increase in age at marriage on an average because of higher exposure to HSAA. We argue longer exposure to property rights reform can increase the women's say in the natal household, including in matters like marriage timing. Further, exposure to reform can also show a positive change in the sticky social norms over time in favor of a woman that can have a bearing in her age of marriage. In fact, Deinenger et al. (2013) found positive effects of HSAA on the age of marriage for the women exposed to reform by almost half a year. To gauge the heterogeneous effect of the exposure on potential changes in age of marriage, we first regress that on *exposure* and then examine whether it can act as a channel to greater empowerment for the women exposed to the reform.

Column 1 of Table 9 indicates that each additional year of *exposure* increases the age at marriage by 0.01 standard deviations, and this is statistically significant at 1% level. This observed gains in age at marriage with exposure to HSAA is after controlling for the education of husband and wife. The result supports our conjecture that norms related to marriage practices might have significantly changed over time in the reform states compared to non-reform states. Next to examine the heterogeneous effect, we interact *exposure* variable with *high_age* dummy that takes the value one for women with positive z-score of age at marriage and re-estimate the models. The coefficients of the interaction variable suggest that the effect of exposure to HSAA on empowerment is higher by 0.006-0.009 standard deviations for women who marry at a higher age (columns 2-4). The higher age at marriage has a greater effect on intrinsic empowerment even though the impact on other domains are also positive and significant at 1% level of significance. Our findings suggest that an increase in age of marriage due to exposure to property rights reform appears to be another channel through which women's ability to make decisions and status in the household improves.

<<Insert Table 9 here>>

### 6.3.3 Change in family structure

Kabeer's (1999) framework suggests that access to resources is necessary, but may not be sufficient for empowering women unless they have control over such resources. The presence of a senior woman in the household, who is likely to have control over these resources, affects the mobility and decision making power of the younger wife in the household (Gram et al., 2019; Anukriti et. al., 2020). Mookerjee (2019) found that implementation of HSAA led to a shift in the family structure from a joint family to a nuclear one which is found to be instrumental in raising female autonomy. If awareness and changing social norms improve over time, it is possible that higher exposure to reforms can lead to a downward shift in the prevalence of joint families. This in-turn, can serve as one of the channels through which we observe improvements in women empowerment via higher exposure to property right reforms. In other words, because of potential inheritance of the ancestral property, the wife and her husband may be more empowered relative to their other family members. Also, women's inheritance rights can increase the younger couple's ability to set up an independent dwelling owning to the higher unearned income of the woman through property rights. Hence the likelihood of them residing in the same household with the wife's in-laws may reduce and therefore in the absence of an elderly, her access to resources may increase and in the process empower her. We study this aspect by examining first if the likelihood of women residing in nuclear families increases because of the property rights reforms and then whether they are systematically more likely to be empowered.

Accordingly to test this, we define joint family as a family structure where couples from two generations reside together under one roof and nuclear family otherwise. Column 1 of Table 10 presents the results highlighting the effects of HSAA on the probability of women residing in a nuclear family. The findings indicate this to be indeed the case, where we find a year increase in *exposure* to HSAA increases the probability of staying in a nuclear family by 0.3 percentage points on an average and this is statistically significant at 1 percent level.

<<Insert Table 10 here>>

Now to examine the heterogeneous effects on women from nuclear families, we interact *exposure* variable with the dummy variable that takes the value of one if woman belongs to a nuclear family and zero otherwise. The marginal effects of this variable in the regression would yield the heterogeneous effect. Indeed as indicated in columns 2-4 of Table 10, we find that the impact of exposure to HSAA on overall empowerment to be higher for women staying in nuclear families relative to those staying in joint families after controlling for age at marriage, husband's and wife's educational attainment. This secular increase seems to be significantly more driven by the increase in instrumental empowerment while no significant effect through the change in family structure from joint to nuclear family is found on intrinsic empowerment. The implications of this suggest that the shift to nuclear families through property rights may help women take important decisions but do not largely change the social norms and perceptions related to women's position within the family.

## 7. Concluding Observations

Studies have argued how the effects of reforms on legal rights to females on the ancestral property are dependent on the awareness levels and also on how entrenched patrilineal social norms are. Given this, the time duration of exposure not only can raise awareness but also can potentially relax the rigid norms and beliefs that work against the ability of the women to exercise her legal rights. In this paper, we examine the effects of duration of exposure of the HSAA implementation in five states of India on a set of indicators of women autonomy that encompass the intrinsic as well as instrumental empowerment. Using the exogenous variation in the implementation of the reforms in these five states, we find that duration of exposure has significant positive effects on both, instrumental and intrinsic empowerment, though the effect size of the former is found to be higher. Our results suggest that an increase in empowerment might have been driven by changes in social norms that are potentially reflected through increased investment in human capital, age at marriage and changes in the family structure to nuclear units.

Notably, when we looked at the heterogeneous effects on caste groups, the gain in empowerment is not uniform as eligible married women from scheduled tribes don't experience any improvements in empowerment that others receive from dissemination effect of HSAA over time. On exploring further, as one would expect, for ST women, higher exposure to HSAA is found to have failed to improve their age at marriage, educational attainment or residence in a nuclear family post marriage.[10] In a way, such exclusionary results bring into fore the role of social norms over and above the legal covenants in Indian society, albeit from different perspectives. For scheduled tribes, the conflict between customary and statutory laws poses obstacles towards effective realization of HSAA benefits. Customary laws of tribes are often discriminatory and divest women from property rights. Except for sporadic cases like Himachal Pradesh High Court ruling in 2015[11] and Mumbai High Court ruling in 2019[12] granting tribal women equal rights under HSAA, such enabling policies have not percolated to other parts of India like Jharkhand. However, even in northern and western states, customs like "*haq tyag*" often forces the daughter to renounce her share in the parental property[13]. Relinquishing legal rights to property is seen as compensation to the parental marriage expenditure that may include dowry payments for daughters. However, the fact that we observe secular effects on empowerment even after controlling for region-specific characteristics indicate that the duration of HSAA might have weakened the effect of such norms. The same cannot be said about the customary tribal laws that not only varies spatially but also across more than 700 tribes and ethnic groups in India. Simultaneously the fact that intrinsic empowerment shows relatively lesser improvement for higher exposure to HSAA indicates that patriarchal norms are still entrenched in the family culture. The fact that newly-wed bride experience lesser empowerment gains from HSAA also lends support to this surmise.

---

[10] The results of the impact of HSAA on age at marriage, education and family structure for the sub-sample of ST women are given in the Appendix Table A7.

[11] https://www.outlookindia.com/newswire/story/tribal-women-have-the-right-to-inherit-as-per-hindu-succession-act-1956-hc/904104 (accessed on October 1, 2020)

[12] https://economictimes.indiatimes.com/wealth/plan/inheritance-rights-of-women-how-to-protect-them-and-how-succession-laws-vary/articleshow/70407336.cms?from=mdr (accessed on October 1, 2020)

[13] https://in.reuters.com/article/india-landrights-women/as-property-prices-rise-more-indian-women-claim-inheritance-idINKBN1QU0PZ (accessed on October 1, 2020)

However, along with social norms, the other important dimension pertaining to legal rights reforms that remains important is the institutional set up. This becomes especially pertinent in India (or global south) that has poor records in court efficiency (Chemin, 2009; Amirapu, 2020). In addition, post implementation supply-side lags that might result from the time taken by the legal practitioners and lengthy legal procedures indicating significant litigation costs especially for those in rural areas might discourage the exercise of right to possession- the attenuating effect on claim might be perpetual. With the duration of exposure, the enabling supply side institutions for implementing the claims under HSAA are likely to get strengthened and indeed this gets reflected from the subsequent amendments including the recent Supreme Court of India ruling on retrospective effect[14]. Due to paucity of data, one main limitation of the paper is we are unable to identify the secular effects of supply-side improvements as channels that may be associated with the passage of HSAA and hence this can qualify as an important future research. Nevertheless, our research identifies the role of the possible improvements in enabling institutions along with relaxation of rigid patrilineal norms and beliefs with the passage of time that potentially can lead to better autonomy for women. Thus, to explore the benefits of HSAA to its fullest potential, policymakers might consider streamlining legal provisions keeping in mind the peculiarities of customs of ethnic groups but at the same time undertake proactive strategies to counter the effects of patriarchy. The design of such policies remains an issue that future research also should look to explore.

---

[14] https://www.livemint.com/news/india/supreme-court-says-daughters-have-equal-property-rights-hindu-succession-amendment-act-2005-has-retrospective-effect-11597135907339.html (accessed on October 1, 2020)

**Declaration**

*Funding:* Authors declare that they did not receive any funding for carrying out this study.

*Conflicts of interest/Competing interests:* The authors declare that they have no conflict of interest.

*Availability of data and material:* Data used for the study is available on request with permission from the Demographic and Health Survey.

*Code availability:* STATA codes are available and have been submitted for replicating the tables and figures.

**References**

Agarwal B (1994) A field of one's own: Gender and land rights in South Asia. Cambridge: Cambridge University Press

Agarwal B (1998) Widows versus daughters or widows as daughters? Property, land, and economic security in rural India. Modern Asian Studies 32: 1-48

Agarwal B, Anthwal P, Mahesh M (2020) Which women own land in India? Between divergent data sets, measures and laws. GDI Working Paper No 2020-043 Manchester: University of Manchester

Ahmed R, Hyndman-Rizk N (2020) The higher education paradox: towards improving women's empowerment, agency development and labour force participation in Bangladesh. Gender and Education 32: 447-465

Alsop R, Heinsohn N (2005) Measuring empowerment: Structuring analysis and framing indicators. Policy Research Working Paper, No. 3510. Washington, D.C. World Bank

Alsop R, Bertelsen M, Holland, J (2006) Empowerment in practice: From analysis to implementation. Directions in Development series, Washington, D.C. World Bank

Amirapu A (2020) Justice Delayed is Growth Denied: The Effect of Slow Courts on Relationship-Specific Industries in India. Economic Development and Cultural Change. https://doi.org/10.1086/711171

Anukriti S, Herrera-Almanza C, Karra M, Pathak PK (2020) Curse of the Mummy-ji: The influence of mothers-in-law on women in India. American Journal of Agricultural Economics. https://doi.org/10.1111/ajae.12114

Banerjee A, Somanathan R (2007) The political economy of public goods: Some evidence from India. Journal of Development Economics 82: 287-314

Behrman JR, Foster AD, Rosenzweig MR, Vashishtha P (1999) Women's schooling, home teaching, and economic growth. Journal of Political Economy 107: 682-714

Behrman JR, King EM (2008) Program impact and variation in duration of exposure. In: Amin S, Das JJ, Goldstein M, (eds) Are You Being Served: New Tools for Measuring Service Delivery. World Bank, Washington, D.C. pp 147–172

Besley T, Leight J, Pande R, Rao V (2016) Long-run impacts of land regulation: Evidence from tenancy reform in India. Journal of Development Economics 118: 72-87

Bose N, Das S (2017) Women's inheritance rights, household allocation, and gender bias. American Economic Review 107:150-153

Chowdhry P (ed) 2017 Women's land rights: Gender discrimination in ownership (Land Reforms in India series). New Delhi: Sage Publications

Caldwell JC, Reddy PH, Caldwell P (1983) Causes of marriage change in south India, Population Studies 37:343-361

Chemin M (2009) Do judiciaries matter for development? Evidence from India. Journal of Comparative Economics *37*: 230-250

Deaton A, Dreze J (2002) A re-examination of poverty and inequality in India. Economic and Political Weekly 37: 3729-3748

Deininger K, Goyal A, Nagarajan H (2013) Women's inheritance rights and intergenerational transmission of resources in India. Journal of Human Resources 48: 114–141

Deininger K, Jin S, Nagarajan H, Xia F (2018) Inheritance law reform, empowerment, and human capital accumulation: Second-generation effects from India. Journal of Development Studies 55: 2549-2571

Desai S, Andrist L (2010) Gender scripts and age at marriage in India. Demography 47: 667–687

Deshpande A (2000) Does caste still define disparity? A look at inequality in Kerala, India. American Economic Review 90: 322-325

Gram L, Morrison J, Skordis-Worrall J (2019) Organising concepts of 'women's empowerment' for measurement: A typology. Social Indicators Research 143:1349–1376

Hahn Y, Nuzhat K, Yang H (2018) The effect of female education on marital matches and child health in Bangladesh. Journal of Population Economics 31: 915–936

Hallward-Driemeier M, Hasan T (2012) Empowering Women: Legal Rights and Economic Opportunities in Africa. Africa Development Forum, Washington, DC, World Bank. https://openknowledge.worldbank.org/handle/10986/11960

Heath R, Tan X (2020). Intrahousehold bargaining, female autonomy, and labor supply: Theory and evidence from India. Journal of the European Economic Association 18: 1928-1968

Hnatkovska V, Lahiri A, Paul SB (2013) Breaking the caste barrier: Intergenerational mobility in India. Journal of Human Resources 48: 435-473

Ibrahim S, Alkire S (2007) Agency and empowerment: A proposal for internationally comparable indicators. Oxford Development Studies 35: 379-403

Jayaweera S (1997) Women, education and empowerment in Asia. Gender and Education 9: 411-424

Jensen R, Thornton, R (2003) Early female marriage in the developing world. Gender and Development 11: 9-19

Kabeer N (1999) Resources, agency, achievements: Reflections on the measurement of women's empowerment. Development and Change 30: 435–464

Kabeer N (2001) Resources, agency, achievements: Reflections on the measurement of women's empowerment. Discussing Women's Empowerment-Theory and Practice. SIDA Studies No. 3, Swedish International Development Cooperation Agency, Sweden.

Kelkar G (2014) The fog of entitlement: Women's inheritance and land rights. Economic and Political Weekly 49: 51-58

Khamis M, Prakash N, Siddique Z (2012) Consumption and social identity: Evidence from India. Journal of Economic Behavior and Organization 83: 353-371

King EM, Behrman, JR (2009) Timing and duration of exposure in evaluations of social programs. World Bank Research Observer 24: 55–82

Mackie G, Moneti F, Shakya H, Denny E (2015) What are social norms? How are they measured. University of California at San Diego-UNICEF Working Paper, San Diego

Maertens A (2013) Social norms and aspirations: Age at marriage and education in rural India. World Development 47: 1-15

Maharatna A (2000) Fertility, mortality and gender bias among tribal population: An Indian perspective, Social Science and Medicine 50: 1333-1351

Maharatna A, Hawley JS (2005) Demographic perspectives on India's tribes. Oxford University Press

Malapit H, Quisumbing A, Meinzen-Dick R, Seymour G, Martinez EM, Heckert J, Rubin D, Vaz A, Yount KM (2019) Development of the project-level Women's Empowerment in Agriculture Index (pro-WEAI). World Development, 122: 675-692

Mare RD (1991) Five decades of educational assortative mating, American Sociological Review, 56: 15-32

Meinzen-Dick R, Quisumbing A, Doss C, Theis S (2019) Women's land rights as a pathway to poverty reduction: Framework and review of available evidence. Agricultural Systems 172: 72-82

Mitra A (2008) The status of women among the scheduled tribes in India. Journal of Socio-Economics 37: 1202-1217

Mookerjee S (2019). Gender-neutral inheritance laws, family structure, and women's status in India World Bank Economic Review 33: 498-515

Mosedale S (2005) Assessing women's empowerment: towards a conceptual framework. Journal of International Development 17: 243-257

Munoz Boudet AM, Petesch P, Turk C, Thumala A (2012) On norms and agency: Conversations about gender equality with women and men in 20 countries. Washington DC: World Bank

Munshi K, Rosenzweig M (2009) Why is mobility in India so low? Social insurance, inequality, and growth (No. w14850). National Bureau of Economic Research

Munshi K (2016) Caste networks in the modern Indian economy. In Development in India. Springer, New Delhi, pp13-37

Narayan-Parker D (2002) Empowerment and Poverty Reduction. Washington DC, World Bank

North D (2005) Understanding the Process of Economic Change. Princeton, Oxford Princeton University Press. doi:10.2307/j.ctt7zvbxt

Petesch P, Smulovitz C, Walton M (2005) Evaluating empowerment: A framework with cases from Latin America. In Narayan-Parker D (ed) Measuring Empowerment: Cross-Disciplinary Perspectives. Washington, DC, World Bank Press, pp 39-68

Premi MK (1990) India. In Nam CB, Serow WJ, Sly DF (eds.) International Handbook on Internal Migration. New York: Greenwood Press

Richardson RA (2018) Measuring women's empowerment: A critical review of current practices and recommendations for researchers. Social Indicators Research 137: 1–19

Roy S (2015) Empowering women? Inheritance rights, female education and dowry payments in India. Journal of Development Economics 114: 233–251

Roychowdhury P (2019) Peer effects in consumption in India: An instrumental variables approach using negative idiosyncratic shocks. World Development 114: 122-137

Samari G (2017) First birth and the trajectory of women's empowerment in Egypt, BMC Pregnancy and Childbirth 17:362. https://doi.org/10.1186/s12884-017-1494-2

Sen A (1985) Commodities and Capabilities. Amsterdam, North-Holland

Sen A (1999) Development as Freedom. New York, Knopf

Singh DP (1998) Internal Migration in India: 1961-1991. Demography India 2: 245-261

Sircar AK, Pal S (2014) What is preventing women from inheriting land? A study of the implementation of the Hindu Succession (Amendment) Act 2005 in three states in India, presented at the 2014 World Bank Conference on Land and Poverty. https://cdn.landesa.org/wp-content/uploads/What-is-Preventing-Women-from-Inheriting-Land-Sircar-Pal-March-2014.pdf

Stark O, Taylor JE (1991) Migration incentives, migration types: The role of relative deprivation. Economic Journal 101: 1163-1178

Stockl H, March L, Pallitto C, Garcia-Moreno C (2014) Intimate partner violence among adolescents and young women: prevalence and associated factors in nine countries: a cross-sectional study. BMC Public Health, 14, 751. https://doi.org/10.1186/1471-2458-14-751

Yount KM, Crandall AA, Cheong YF (2018) Women's age at first marriage and long-term economic empowerment in Egypt. World Development 102: 124-134